# Brightness Characterization and Modeling for Amazon Leo Satellites

Anthony Mallama [1], Richard E. Cole [1], Hui Zhi [1], Brad Young [1],
Jay Respler [1], Olga Zamora [1,2] and Michelle Dadighat [1,3]

[1] IAU Centre for the Protection of the Dark and Quiet Sky
[2] Instituto de Astrofísica de Canarias and Universidad de La Laguna
[3] National Science Foundation, NOIRLab

2026 January 12

Correspondence: anthony.mallama@gmail.com

The mean apparent magnitude of Amazon Leo satellites is 6.28 based on 1,938 observations. For spacecraft in their operational mode, 92% exceeded the brightness limit recommended by the IAU for interference with research, while 25% distract from aesthetic appreciation of the night sky. The reflective characteristics are similar to Version 1 Starlink spacecraft.

## 1. Introduction

Bright spacecraft interfere with astronomical observations (Barentine et al. 2023, and Mallama and Young 2021). Satellites are luminous in the visible part of the electromagnetic spectrum due to the reflection of sunlight and at radio wavelengths because they transmit signals.

The Amazon Leo constellation (formerly named Project Kuiper) concerns astronomers because 3,232 satellites are planned (FCC, 2024). Satellite attitude determines how sunlight is reflected and influences brightness. This paper discusses the attitude control modes in use and reports on spacecraft brightness.

Section 2 describes the orbits of Amazon Leo satellites and their evolution. Section 3 discusses how brightness measurements were obtained for this study. Section 4 characterizes the magnitudes of the satellites from an empirical perspective. Section 5 explains those brightness characteristics using a physical model. Section 6 discusses the impact of these spacecraft on astronomy. Section 7 summarizes our conclusions.

## 2. Orbits and orbit-raising

Amazon launched the first 27 satellites aboard an Atlas V rocket on 2025 April 28. The spacecraft were injected into 51.9° inclination orbits at 480 km and then ascended to 630 km over a period of several months as shown in Figure 1.

Spacecraft may be quite bright during orbit-raising because their mean distances to observers are smaller than when they are on station. Furthermore, satellite operators do not always implement brightness mitigation attitudes during this period.

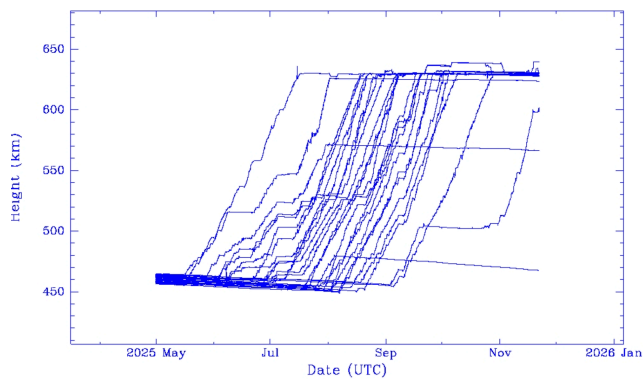

*Figure 1. Orbit-raising for the first launch of Amazon Leo satellites. Credit: Jonathan McDowell, https://planet4589.org.*

Six subsequent launches on Atlas and Falcon 9 rockets have placed more spacecraft into similar orbits as indicated by Table 1.

Table 1. Amazon Leo launches

| Date of 2025 | Rocket | Satellites |
|---|---|---|
| April 28 | Atlas V 551 | 27 |
| June 23 | Atlas V 551 | 27 |
| July 16 | Falcon 9 | 24 |
| August 11 | Falcon 9 | 24 |
| September 25 | Atlas V 551 | 27 |
| October 14 | Falcon 9 | 24 |
| December 16 | Atlas V 551 | 27 |

Additional launch vehicles will be used in 2026 as listed in Figure 2. They will orbit greater numbers of satellites per launch. Amazon has stated that later spacecraft will operate at heights of 610 and 590 km.

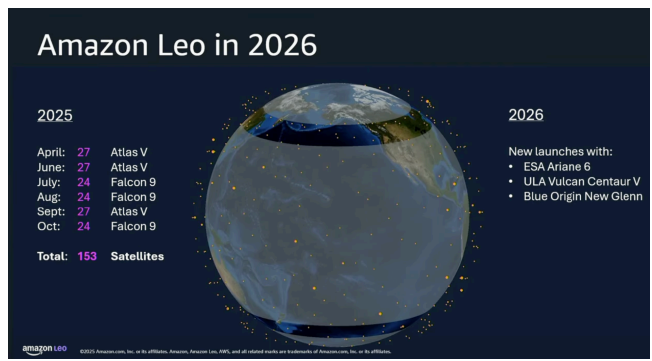

*Figure 2. Current and future launch vehicles. Credit: Amazon.*

## 3. Observations

Satellite magnitudes were measured using detector-based and visual techniques. Electronic measurements were recorded by the MMT9 robotic observatory in Russia (Karpov et al. 2015 and Beskin et al. 2017) located at 43.65N and 43.65E. The magnitudes are within 0.1 of the Johnson V-band as discussed by Mallama (2021). Data was recorded at a frequency of 10 Hz which we averaged into 5 second means.

Additional electronic magnitudes were recorded with the 50 cm telescope at the National Astronomical Observatories in China at 40.393N 117.574E. These measurements of orbit-raising satellites were taken in the Gaia G-band which approximates the V-band although it is wider.

Visual magnitudes were determined by comparing the spacecraft to nearby reference stars. The angular proximity between satellites and stellar objects accounts for variations in sky transparency and sky brightness. Mallama (2022) describes this method in more detail. Many of the visual magnitudes fill gaps in the Sun-satellite-observer geometry that were not recorded electronically. Visual observers were located in the UK at 50.55N 4.73W and in the US at 38.98N 76.76W, 36.13N 95.99W and 40.33N 74.44W.

## 4. Brightness characterization

The impact of satellites on astronomy is directly related to their luminosity, which highlights the importance of brightness characterization. Table 2 indicates that the mean of all apparent magnitudes for all measurements of the Amazon Leo constellation is 6.28 while its standard deviation (SD) is 0.96 and the standard error of the mean (SEM) is 0.02. Once satellites reach their operational altitude, the mean, SD and SEM for magnitudes are 6.43, 0.90 and 0.03. That mean is fainter than the value for all observations because spacecraft at 630 km are typically at greater distances from the observer.

Table 2. Magnitude statistics.

|  | Apparent magnitude | | | 1000-km magnitude | | | |
|---|---|---|---|---|---|---|---|
|  | Mean | SD | SEM | Mean | SD | SEM | # Obs |
| Overall | 6.28 | 0.96 | 0.02 | 6.93 | 0.94 | 0.02 | 1938 |
| Operational Height | 6.43 | 0.90 | 0.03 | 6.83 | 0.91 | 0.03 | 1086 |
| Below Operational Height. | 6.10 | 0.99 | 0.03 | 7.05 | 0.97 | 0.03 | 852 |

Apparent magnitudes may be adjusted to a uniform range of 1,000 km to remove the effect of distance. Table 2 shows that the mean, SD and SEM of all magnitudes adjusted to 1000-km are 6.93, 0.94 and 0.02. The corresponding values for magnitudes at operational heights are 6.83, 0.91 and 0.03. That mean value is brighter than the mean for orbit-raising satellites. Magnitude statistics pertaining to the impact of Amazon satellites on astronomy are discussed in Section 6.

## 5. Physical Model

As with other constellations, the spacecraft brightness before and during ascent to their final orbit is not considered as critical as its brightness in operations. Therefore, the analysis in this paper needed to separately analyze spacecraft observations made under those two conditions, termed here 'orbit-raising' and 'operational'.

Amazon has released a small number of images and videos of the Leo spacecraft during deployment and on-orbit. Figure 3 indicates there are mirror surfaces on the spacecraft.

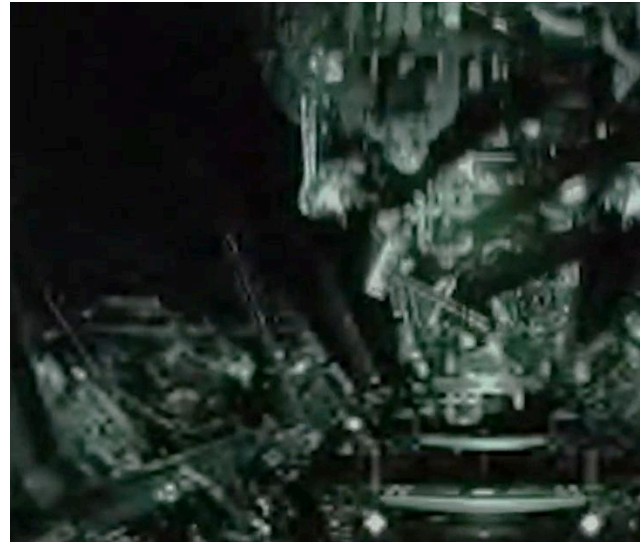

Figure 3. Image of an Amazon Leo spacecraft (right) deploying from its Atlas-V launcher. Reflection from a mirror surface on the adjacent spacecraft can be seen (left) Credit: Amazon.

Figure 4 shows a possible configuration on the Amazon LEO spacecraft, developed by the authors based on the available images and videos available as of November 2025. Our interpretation is that three antenna panels are folded at launch and deploy to form a large, nadir-facing panel, at least part of which is mirror-faced (reflections can be seen in Figure 3). A number of auxiliary equipment items are visible in the images, mounted on the central section. The videos suggest the solar panel can rotate around its long axis to track the Sun. This configuration cannot be completely accurate but is used as a guide in the interpretation that follows.

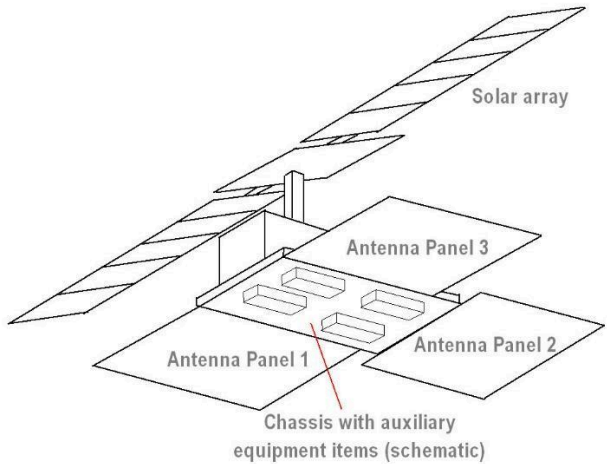

*Figure 4. A possible configuration of the Amazon LEO spacecraft.*

The height of the Amazon Leo spacecraft can be derived from publicly available orbital elements but Amazon has not published the detailed design of the spacecraft or its attitude control modes once in orbit. Therefore, a method of distinguishing spacecraft attitude mode by observation was required to properly characterize their brightnesses.

Observations taken of the first Amazon Leo spacecraft during the first few months after launch showed a wide range of magnitudes during a single pass, from binocular range to easy naked-eye. A simple optical model of a single diffusely-reflecting surface was used to investigate the pointing direction of the main reflecting surface, adjusting the modeled orientation to match predictions to the observation set. It was clear that the reflecting surface was not directly sun-pointing but was oriented in, or close to, the plane of the orbit, at least at the times the spacecraft were observed (Figure 5). From their nature, observations are made when the Sun is close to the horizon at the spacecraft and below the horizon at the observer.

Depending on the orbit beta angle (the angle between the plane of the orbit and the direction of the Sun) at the time of observation, the angle of incidence of sunlight on the modeled reflecting surface could vary between 0 and 90°. A correction was made to the apparent magnitude to account for that variability (assuming diffuse reflection) and for the distance of the spacecraft, using the standard of 1000km. A plot of these corrected magnitudes against the calculated angle of reflection from the panel to the observer is shown in Figure 6. The objects are bright when the angle of reflection is small and faintest when that angle is greater than 90°, where the sunlit side of the panel is turned away from the observer.

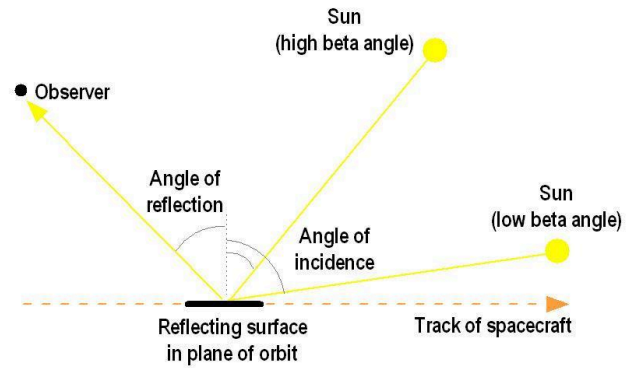

*Figure 5: Simplified plan view of the satellite track with the reflecting surface in the plane of the orbit (orbit-raising mode only). The Sun position and angles are shown for cases of low and high orbit beta angle.*

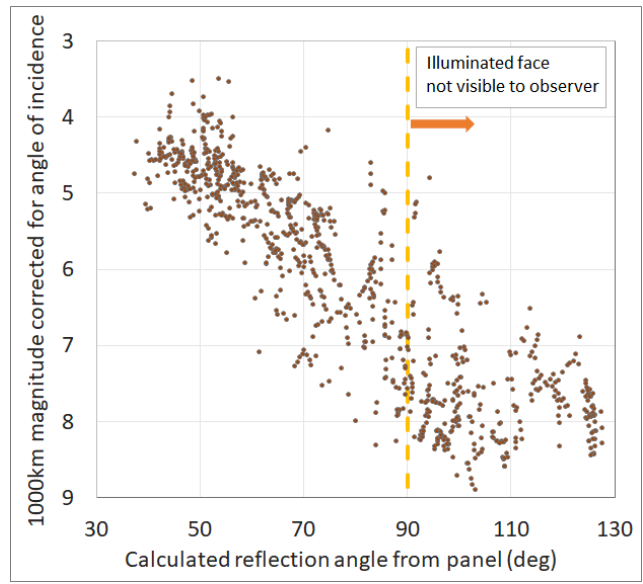

*Figure 6. Observed magnitudes of Amazon Leo objects in orbit-raising mode, corrected for range and angle of sunlight incidence, plotted against angle of reflection to the observer.*

Amazon later published a video taken from one of their spacecraft soon after launch when it would likely be in orbit-raising mode, including some details of its track across the Earth (Figure 7). The solar panel is seen in the image; oriented close to the local vertical when the spacecraft was close to eclipse, according to the track stated by Amazon.

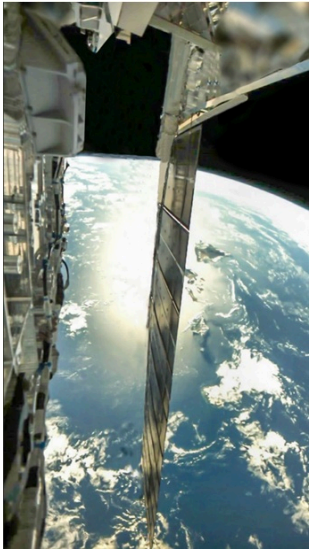

*Figure 7. Image taken from a video of an Amazon Leo spacecraft in early-orbit operations. It shows the spacecraft in orbit-raising mode. Credit: Amazon.*

On this basis, observations that showed the optical behavior in Figure 6 were graded as objects in the orbit-raising mode.

Later observations of Amazon Leo objects in their final operational orbit showed many (but not all) in a different attitude configuration ("operational") where the objects were bright at low phase angle, fainter at zenith and then sometimes bright at high phase angle. This behavior appears similar to the first generation of Starlink spacecraft which had a similar basic configuration to Amazon Leo, that is an approximately sun-facing solar panel and a partly mirror-surfaced nadir panel (Cole 2021).

Thirty-nine different spacecraft were observed in operational mode and 65 in orbit-raising mode.

Figure 8 shows a plot of 1000 km magnitude against phase angle for the Amazon Leo observations graded as objects in operational mode. For a directly sun-facing surface, phase angle is the same quantity as the angle of reflection from that panel to the observer.

The spacecraft is bright at low phase angles due to reflection from an approximately sun-facing surface, presumably the solar panel.

The spacecraft is bright at high phase angles due to specular reflection by the mirror-faced nadir panel of sun-illuminated earth surface, as occurred in the first generation Starlink and the Gen2 mini Starlink (Fankhauser et al 2023).

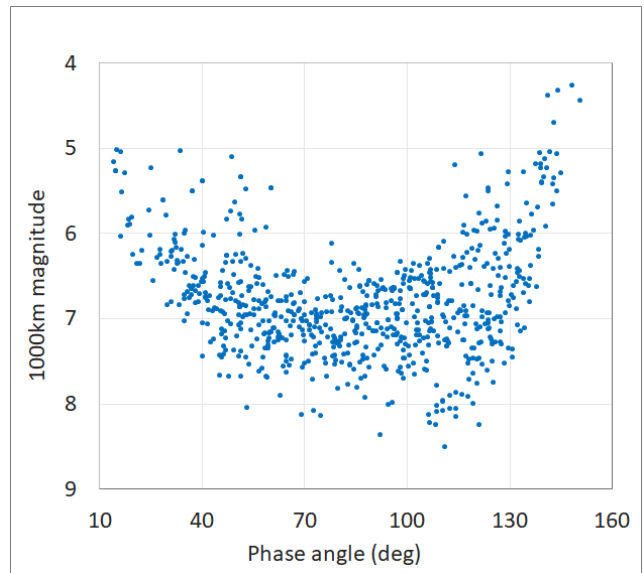

*Figure 8. Observations of Amazon Leo spacecraft in operational mode at operational height.*

The elevation angle of the Sun at the geographical point reflected by a nadir-facing mirror to the observer was calculated for the Amazon Leo operational mode observation set. Figure 9 plots the measured 1000 km magnitudes against Sun elevation angle and indicates that the brightness increases when the Sun is above an elevation of -5° at the reflected geographical locations. That brightening indicates that the satellites are reflecting light from the Earth's dayside, increasing their brightness by over two magnitudes, up to visual magnitude 5.5 at elevations below 25°.

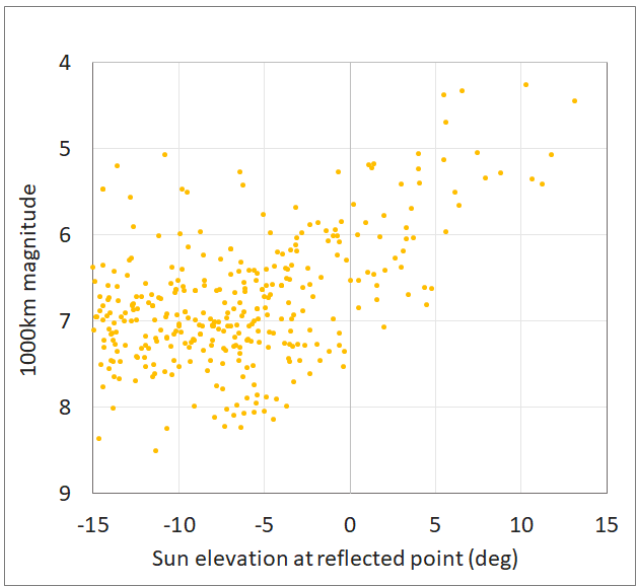

*Figure 9. Plot of 1000 km brightness of Amazon Leo objects in operational mode against the Sun elevation angle at the geographical point specularly reflected to the observer by the nadir panel. The objects are brighter when the Sun is above the horizon.*

## 6. Impact on astronomy

The International Astronomical Union (IAU, 2024) recommended an *acceptable brightness limit* which states that satellites in operational orbits should not be visible to the unaided eye. Objects of visual magnitude 6 can be seen at locations where the sky is minimally affected by light pollution. We refer to this magnitude as the *aesthetic limit*.

The IAU statement also defined a brightness limit for interference with professional astronomy which we call the *research limit*. For altitudes up to 550 km that limit is magnitude 7.0. Equation 1 specifies the limit for altitudes above 550 km,

$M_V > 7.0 + 2.5 * \log_{10}(altitude / 550)$

Equation 1

where $M_V$ is the Johnson visual magnitude and *altitude* refers to the satellite's height above sea level in km. Equation 1 evaluates to $M_V$ = 7.15 at 630 km.

Table 3 lists magnitude statistics for satellites maintaining a height of 630km, in operational mode or still in the attitude for orbit-raising, based on the distinction explained in Section 5. The means are nearly identical although the standard deviations of operational spacecraft magnitude are less than half as large. This difference in standard deviation is due to the smaller variability in illuminated spacecraft surface visible to the observer, in operational attitude.

Table 3. Magnitude statistics for satellites maintaining a height of 630 km

| | Apparent magnitude | | | 1000-km magnitude | | | |
|---|---|---|---|---|---|---|---|
| Control Mode | Mean | SD | SEM | Mean | SD | SEM | # Obs |
| Operational | 6.44 | 0.62 | 0.02 | 6.81 | 0.64 | 0.02 | 758 |
| Orbit-raising | 6.41 | 1.34 | 0.07 | 6.87 | 1.33 | 0.07 | 328 |

Figure 10 compares the brightness distribution of satellites at 630 km to the IAU recommended limits. For satellites in their operational mode, 92.0% of observations exceed the research limit, while 24.7% exceed the aesthetic limit.

Figure 11 compares the brightness of operational Amazon Leo satellites with other satellite constellations. The closest in luminosity are Starlink spacecraft in 485 km orbits.

The analysis in this paper pertains to Amazon Leo satellites at 630 km altitude. Future spacecraft which are expected to orbit at lower altitudes will be brighter. Amazon Leo has recently stated that it is deploying ground-based observatories to measure reflectivity brightness with the intent to improve performance through block upgrades.

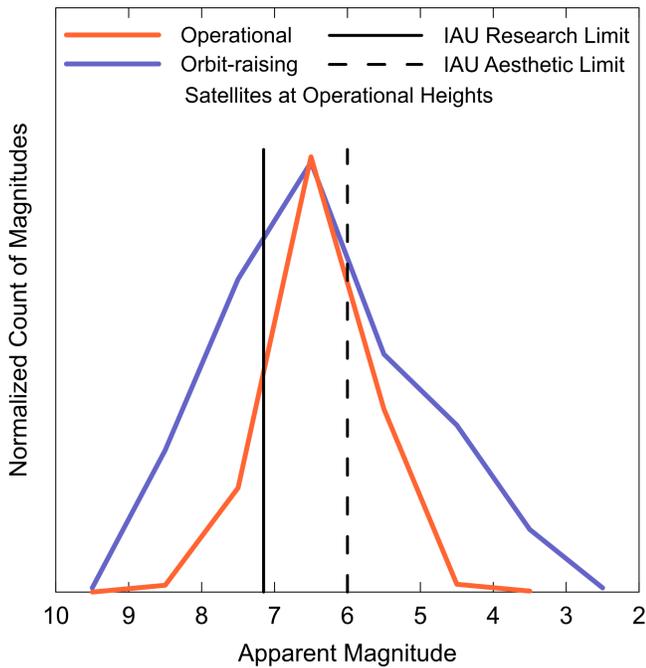

*Figure 10. The distribution of apparent magnitudes for satellites in operational and orbit-raising modes at 630 km is compared to the research and aesthetic limits.*

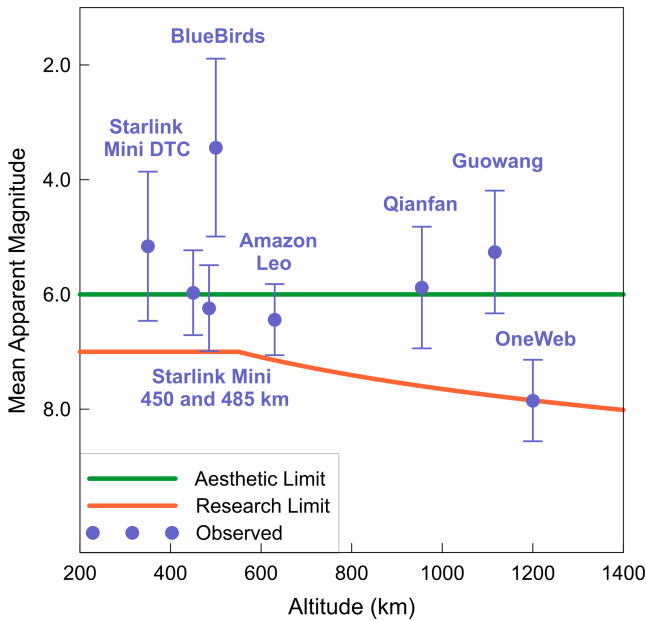

*Figure 11. The mean brightness and its standard deviation for Amazon Leo satellites in their operational mode are compared to other constellations as reported by Mallama & Cole 2025.*

## 7. Conclusions

The mean apparent magnitude of all Amazon Leo satellites is 6.28 based on 1,938 observations. For spacecraft in their operational mode, 92% exceeded the IAU brightness limit for interference with research, while 25% distract from aesthetic appreciation of the night sky.

The reflective characteristics are similar to Version 1 Starlink spacecraft. Both strongly scatter sunlight forward and backward. Based on private communication, Amazon is working on reducing satellite brightness.

The Amazon Leo constellation spacecraft are potentially impacting astronomical research and aesthetic appreciation of the night sky.

Brightness statistics for all major satellite constellations are kept up to date at our website, *https://satmags.netlify.app/*.

**Acknowledgments**
We thank C. Hofer of the Amazon Leo project and an anonymous reviewer for their valuable comments on an earlier version of this paper.

We also thank E. Katkova for maintaining a public database of MMT9 satellite observations and S. Karpov for his correspondence with us about MMT9.

The website of Heavens-Above was used to predict satellite passes. The planetarium program, Stellarium, Heavensat and the IAU/CPS SatChecker app were used for data analysis.

This research made use of data and/or services provided by the International Astronomical Union Centre for the Protection of the Dark and Quiet Sky (IAU CPS) SatHub. The development of SCORE has been supported by the National Science Foundation (NSF) under grant number: AST 2332736. SCORE is hosted at the NSF NOIRLab. Any opinions, findings, and conclusions or recommendations expressed in this material are those of the author(s) and do not necessarily reflect the views of the NSF NOIRLab, the SKAO, the IAU, or any host or member institution of the IAU CPS.

**Data availability**

The observations used in this study can be accessed from the IAU/CPS SCORE database at https://score.cps.iau.org/.